\documentclass[aos]{imsart}

\RequirePackage[OT1]{fontenc}
\RequirePackage{amsthm,amsmath,natbib}
\RequirePackage[colorlinks]{hyperref} \RequirePackage{hypernat}
\usepackage{amsthm}
\usepackage{amsbsy}
\usepackage{epic}
\usepackage{graphicx}
\usepackage{graphics}
\usepackage{amssymb}
\usepackage{psfrag}
\usepackage{amsmath}
\usepackage{epsfig}
\usepackage{array}

\newtheorem{theorem}{Theorem}[section]

\newtheorem{proposition}[theorem]{Proposition}
\newtheorem{definition}[theorem]{Definition}
\newtheorem{lemma}[theorem]{Lemma}

\def\ifdraft{\ifx34}

\def\proof{\paragraph{\sc Proof}\quad}

\def\sq{\hbox{\rlap{$\sqcap$}$\sqcup$}}
\def\qed{\ifmmode\sq\else{\unskip\nobreak\hfil\penalty50\hskip1em\null\nobreak\hfil\sq\parfillskip=0pt\finalhyphendemerits=0\endgraf}\fi}

\def\eqdef{\mathbin{:=}}
\def\eq#1/{(\ref{e:#1})}

\def\beq{\begin{equation}}
\def\eeq{\end{equation}}
\newcommand\ba{\begin{array}{rcl}}
\newcommand\ea{\end{array}}

\def\beqa{\begin{eqnarray}}
\def\eeqa{\end{eqnarray}}
\newcommand{\et}{\end{theorem}}
\newcommand{\bt}{\begin{theorem}}

\def\N{\mbox{N}}

\def\Ga{\mbox{Ga}}

\def\Ca{\mbox{Ca}}

\newcommand{\comment}[1]{}

\def\bX{{\bf X}}
\def\bY{{\bf Y}}
\def\b1{{\bf 1}}
\def\bZ{{\bf Z}}
\def\bz{{\bf 0}}

\def\Pg{{\cal P}}
\def\Pone{{\cal P} _1}
\def\P0{{\cal P}_0}

\def\tX{{\tilde X}}

\def\L{{\cal L}}

\def\Prob{{\bf P}}
\def\E{{\bf E}}
\def\sx{\sigma _2}
\def\sy{\sigma _1}

\def\archive#1{}

\begin{document}

\begin{frontmatter}

\title{STABILITY OF THE GIBBS SAMPLER FOR BAYESIAN HIERARCHICAL
MODELS\protect} \runtitle{STABILITY OF GIBBS SAMPLER}

\begin{aug}
  \author{\fnms{Omiros} \snm{Papaspiliopoulos}\thanksref{t2} \ead[label=e1]{o.papaspiliopoulos@warwick.ac.uk}
  \ead[label=u2,url]{http://www.warwick.ac.uk/staff/O.Papaspiliopoulos/}  }
  \and
  \author{\fnms{Gareth} \snm{Roberts}\ead[label=e2]{g.o.roberts@lancaster.ac.uk}
  \ead[label=u3,url]{http://www.maths.lancs.ac.uk/~robertgo/} }

  \thankstext{t2}{Research funded by EPSRC grant GR/S61577/01}

  \runauthor{Papaspiliopoulos and Roberts}

  \affiliation{University of Warwick and Lancaster University}

  \address{
Mathematics Institute\\
University of Warwick\\
Coventry, CV4 7AL\\
\printead{e1}\\
}

  \address{
Department of Mathematics \& Statistics\\
Lancaster University\\
Lancaster, LA1 4YF\\
\printead{e2} \\
}

\end{aug}

\begin{abstract}
We characterise the convergence of the Gibbs sampler which samples
from the joint posterior distribution of parameters and missing
data in  hierarchical linear models with arbitrary symmetric error
distributions. We show that the convergence can be uniform,
geometric or sub-geometric depending on the relative tail
behaviour of the error distributions, and on the parametrisation
chosen. Our theory is applied to characterise the convergence of
the Gibbs sampler on latent Gaussian process models. We indicate
how the theoretical framework we introduce will be useful in
analyzing more complex models.
\end{abstract}



\end{frontmatter}


 \section{Introduction}\label{sec:intro} Hierarchical modelling
is a widely adopted approach to constructing complex statistical
models. The appeal of the method lies in the simplicity in
specifying a highly multivariate model by joining many simple and
tractable models, the foundational justification based on the
ideas of partial exchangeability, the flexibility to extend or
simplify the model in the light of new information, and the ease
of inference using powerful Markov chain Monte Carlo (MCMC)
methods which have been developed to this end during the last two
decades. Thus, hierarchical models have been used in many areas of
applied statistics such as geostatistics \citep{MR1626544},
longitudinal analysis \citep{Digg:Lian:Zege:1994}, disease mapping
\citep{MR92d:62032}, and financial econometrics \citep{omi} to
name just a few.

A rather general form of a two-level hierarchical model is
\begin{eqnarray}
Y  & \sim & \L(Y | X) \nonumber \\
X &  \sim  &  \L(X | \Theta)\,, \label{e:hier}
\end{eqnarray}
where  $\L(X)$ and $\L(Y \mid X)$ denote the distribution of $X$
and the conditional distribution of $Y$ given $X$ respectively. We
will refer to $Y$ as the data, $X$ as the missing data  and
$\Theta$ as the parameters. In a Bayesian context the model is
completed by specifying a prior distribution for $\Theta$.
Typically the dimension of $X$ is much larger than that of
$\Theta$ and it can increase with the size of the data set. Most
of the applications cited above fit into (\ref{e:hier}) by
imposing the appropriate structure on $\L(Y\mid X)$ and $\L(X \mid
\Theta)$. It is straightforward to construct models with more
levels.

Bayesian inference for (\ref{e:hier}) involves the posterior
distribution $\L( X,\Theta \mid Y=y)$.  This is typically
analytically intractable, but it can be sampled relatively easily
using the Gibbs sampler \citep[][]{MR94g:62056}, by simulating
iteratively from the two conditional distributions $\L(X \mid
\Theta,Y=y)$, and $\L(\Theta \mid X,Y=y)$.  It has been
demonstrated both theoretically and empirically that the
convergence (to be formally defined in Section \ref{results}) of
the Gibbs sampler relates to the structure of the hierarchical
model and particularly to the dependence between the updated
components, $X$ and $\Theta$. Nevertheless, the exact way in which
the model structure interferes with the convergence remains
largely unresolved. Concrete theoretical results exist only for
Gaussian hierarchical models, but we will see that these results
do not extend to more general cases. Although interesting
characterizations of the convergence rate in terms of the
dependence between $X$ and $\Theta$ exist when the Gibbs sampler
is geometrically ergodic \citep{MR92i:60069}, there exist no
general results which establish geometric ergodicity for the Gibbs
sampler. The difficulty in obtaining such general results lies in
the intrinsic dependence of the convergence of the Gibbs sampler
on the model structure.

In this paper we show explicitly how the relative tail behaviour
of $\L(Y \mid X)$ and $\L(X \mid \Theta)$ determines the stability
of the Gibbs sampler, i.e.\@ whether the convergence is uniform,
geometric or sub-geometric. Moreover, we show that the relative
tail behaviour dictates the type of parametrisation that should be
adopted. In order to retain tractability and formulate
interpretable and easy to check conditions we restrict attention
to the class of linear hierarchical models with general error
distributions; the precise model structure is given in Section
\ref{sec:bg}. Nevertheless, our main theoretical results, in
particular Theorems \ref{th:unifnon}, \ref{th:non-g}, \ref{th:geo}
and \ref{th:rwlike}, and the methodology for proving them are
expected to be useful in a much more general context than the one
considered here.

Consideration of the class of linear non-Gaussian hierarchical
models is not merely motivated by mathematical convenience. These
models are very useful in real applications, for example in
longitudinal random effects modelling
\citep{Digg:Lian:Zege:1994,Lair:Ware:rand:1982}, time series
analysis \citep{MR95h:62177,MR95a:62073,MR922169} and spatial
modelling \citep{MR1626544}. They also are a fundamental tool in
the robust Bayesian analysis
\citep[][]{MR49:1662,ohagan:1979,peri:smith:1992,Wake:Smit:Raci:Gelf:baye:1994}.
Furthermore, we will see that the stability of the Gibbs sampler
for linear non-Gaussian models is very different compared to the
Gaussian case, the local dependence between $X$ and $\Theta$ being
crucial in the non-Gaussian case. Notice that several other models
can be approximately written as linear non-Gaussian models.
Actually, this work has been motivated by the behaviour of MCMC
for non-Gaussian Ornstein-Uhlnebeck stochastic volatility models
\citep{omi}.

The paper is organised as follows. Section \ref{sec:bg} specifies
the models we will be concerned with and it establishes some basic
notation. Section \ref{sec:param} discusses Gibbs sampling under
different parametrisations of the model and Section
\ref{sec:example} motivates the theory and the methodology
developed in this paper by a simple example. Section \ref{results}
is the theoretical core of this paper; the section commences with
a short review of stability concepts for the Gibbs sampler;
Section \ref{sec:gs} recalls the existing results for Gaussian
linear models; Section \ref{sec:gen-th} develops stability theory
for hierarchical models and states three main theorems for the
stability of the Gibbs sampler; based on these theorems Section
\ref{sec:comp} provides the characterization of the stability of
the Gibbs sampler under different parametrisations for a broad
class of linear hierarchical models; Section \ref{sec:collapsed}
considers an alternative augmentation scheme when one of the error
distributions is a scale mixture of normals and compares the
convergence of a three-component Gibbs sampler with that of its
collapsed two-component counterpart. Section \ref{sec:latent}
extends the theory to hierarchical models which involve latent
Gaussian processes. Section \ref{sec:discuss} discusses extensions
and contains  some practical guidelines. Section \ref{sec:proofs}
contains the proofs of all theorems and propositions. The proofs
are based on establishing geometric drift conditions and
minorization conditions and using capacitance arguments in
conjunction with Cheeger's inequality.

\section{Models, parametrisations and motivation}

\subsection{Linear hierarchical models} \label{sec:bg}

The models we consider in this paper are of the following form,
where ${\bf Y}_i$ is $m_i \times 1$, ${\bf C}_i$ is $m_i \times
p$, ${\bf X}_i$ is $p \times 1$, ${\bf D}$ is $p \times 1$ and
$\Theta$ is a scalar:
\begin{eqnarray}
{\bf Y}_i  & = & {\bf C}_i {\bf X}_i+{\bf Z}_{1i} \ ,~~~i=1,\ldots,m  \nonumber \\
{\bf X}_i & = &  {\bf D}\Theta +{\bf Z}_{2i} \ . \label{e:random}
\end{eqnarray}
${\bf Z}_{1i},i=1,\ldots,m$, are iid with distribution $\L({\bf
Z}_1)$, ${\bf Z}_{2i},i=1,\ldots,m$, are iid with distribution
$\L({\bf Z}_2)$, and $\L({\bf Z}_1)$ and $\L({\bf Z}_2)$ are
symmetric distributions around {\bf 0} (a vector of 0s with the
appropriate dimension). In the sequel, bold-face letters will
correspond to vectors and matrices, capital letters to random
variables and lower-case letters to their realisations. In this
setting ${\bf Y}=({\bf Y}_1,\ldots,{\bf Y}_m)$ and ${\bf X}=({\bf
X}_1,\ldots,{\bf X}_m)$. The first equation in (\ref{e:random})
will be termed the {\em observation equation} and the second  the
{\em hidden equation}.

 It is
often conveniently assumed that both $\L({\bf Z}_1)$ and $\L({\bf
Z}_2)$ are Gaussian. However there are several applications where
this assumption is clearly inappropriate,  especially if we wish
to make the inference about ${\bf X}$ robust in the presence of
prior-data conflict. It is known \citep[see e.g.][and references
therein]{ohagan:1979,peri:smith:1992,Wake:Smit:Raci:Gelf:baye:1994}
that if the tails of $\L({\bf Z}_1)$ are heavier than the tails of
$\L({\bf Z}_2)$ then inference for ${\bf X}$ is robust  to
outlying observations, whereas if $\L({\bf Z}_2)$ has heavier
tails than $\L({\bf Z}_1)$ inference for ${\bf X}$ is less
influenced by the prior in case of data-prior conflict; these
robustness is absent from Gaussian models. This type of robust
modelling has been undertaken in time-series analysis, see for
example \citep{MR922169}.

\subsection{Gibbs sampling and parametrisations}
\label{sec:param}
As is common in this framework, we place an
improper flat prior on $\Theta$, which in this context leads to a
proper posterior. Bayesian inference for (\ref{e:random}) involves
the joint posterior distribution $\L({\bf X},\Theta \mid {\bf
Y=y})$, which will abbreviate to $\L({\bf X},\Theta \mid {\bf
Y})$.
Although it is often analytically intractable, it can be sampled
easily using the Gibbs sampler.

The parametrisation $\P0 \eqdef ({\bf X}, \Theta )$ is termed the
{\em centred parametrisation}. This terminology was first used in
the linear Gaussian context by \citep{MR1366275}. Following
\citep{MR2003180} we shall use the term more generally to refer to
a parametrisation where the parameters and the data are
conditionally independent given the missing data. We can use the
Gibbs sampler to collect samples from $\L({\bf U},\Theta \mid {\bf
Y})$ where ${\bf U}=h({\bf X},\Theta)$, for some invertible
transformation $h$, and then transform the draws to obtain samples
from $\L({\bf X},\Theta \mid {\bf Y})$. In the rest of the paper
we will use  $\Pg$ to refer to a general parametrisation $({\bf
U},\Theta)$. It is known \citep{MR95d:62133} that the convergence
(to be formally introduced in Section \ref{results}) of the Gibbs
sampler improves as the dependence between the updated components,
${\bf U}$ and $\Theta$, decreases.  Hence, the development of
general re-parametrisation strategies  has been actively
researched, see \citep{MR2003180} for a recent account.  In that
work, the authors introduce the {\em non-centred
  reparametrisation} $\Pone \eqdef ({\bf \tX} ,\Theta )$, which replaces
  ${\bf X}$ with ${\bf \tX} \eqdef h({\bf X},\Theta)$, where $h$
  is a transformation which makes $\Theta $ and
${\bf \tX}$  apriori independent.  In the context of linear
hierarchical models ${\bf \tX} = ({\bf \tX}_1,\ldots,{\bf
\tX}_m)$, where ${\bf \tX}_i=h({\bf X}_i,\Theta)$, and $h({\bf
x},\theta):={\bf x}- {\bf D} \theta $. We will see that $\P0$ and
$\Pone$ present two natural choices.

The prolific expansion in the use of Gibbs sampling for inference
in hierarchical models during the 1990s was  fuelled by the
apparent rapid convergence of the algorithm in many cases.
However, to date, there has been little theoretical analysis
linking the stability of the Gibbs sampler to the structure of
hierarchical models. A notable exception are the explicit
convergence results for  Gaussian linear hierarchical models
obtained in \citep{MR98h:60100} and  summarised in Section
\ref{sec:gs}. The following example is revealing as to what might
go wrong when considering non-Gaussian linear models, and
motivates the methodology and theory developed in this article.

\subsection{ A motivating example} \label{sec:example}

\begin{center}
\begin{figure}[htb]
\begin{tabular}{cc}
\psfig{figure=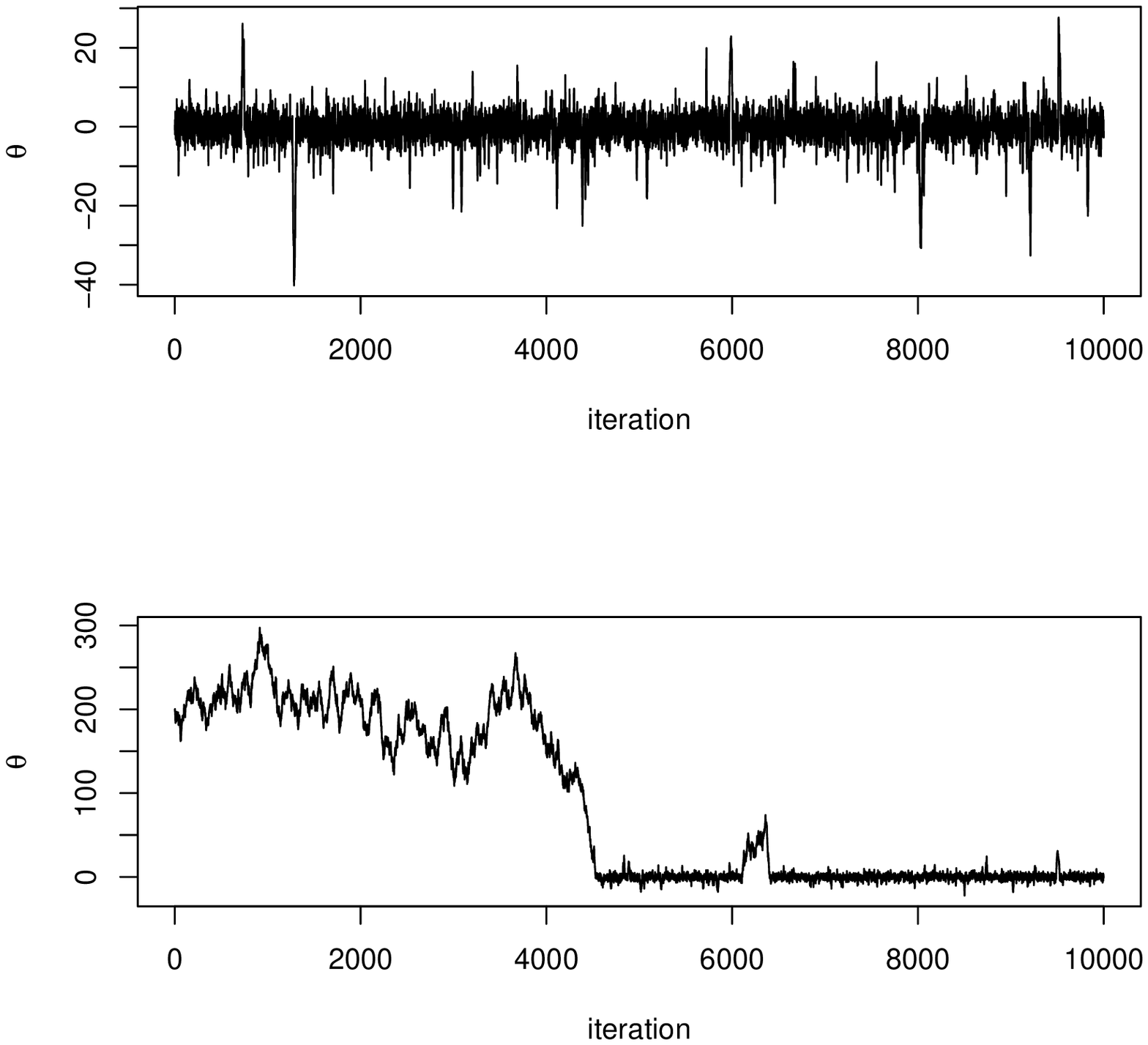,height=2.0in,width=3.0in,  angle=270} &
\psfig{figure=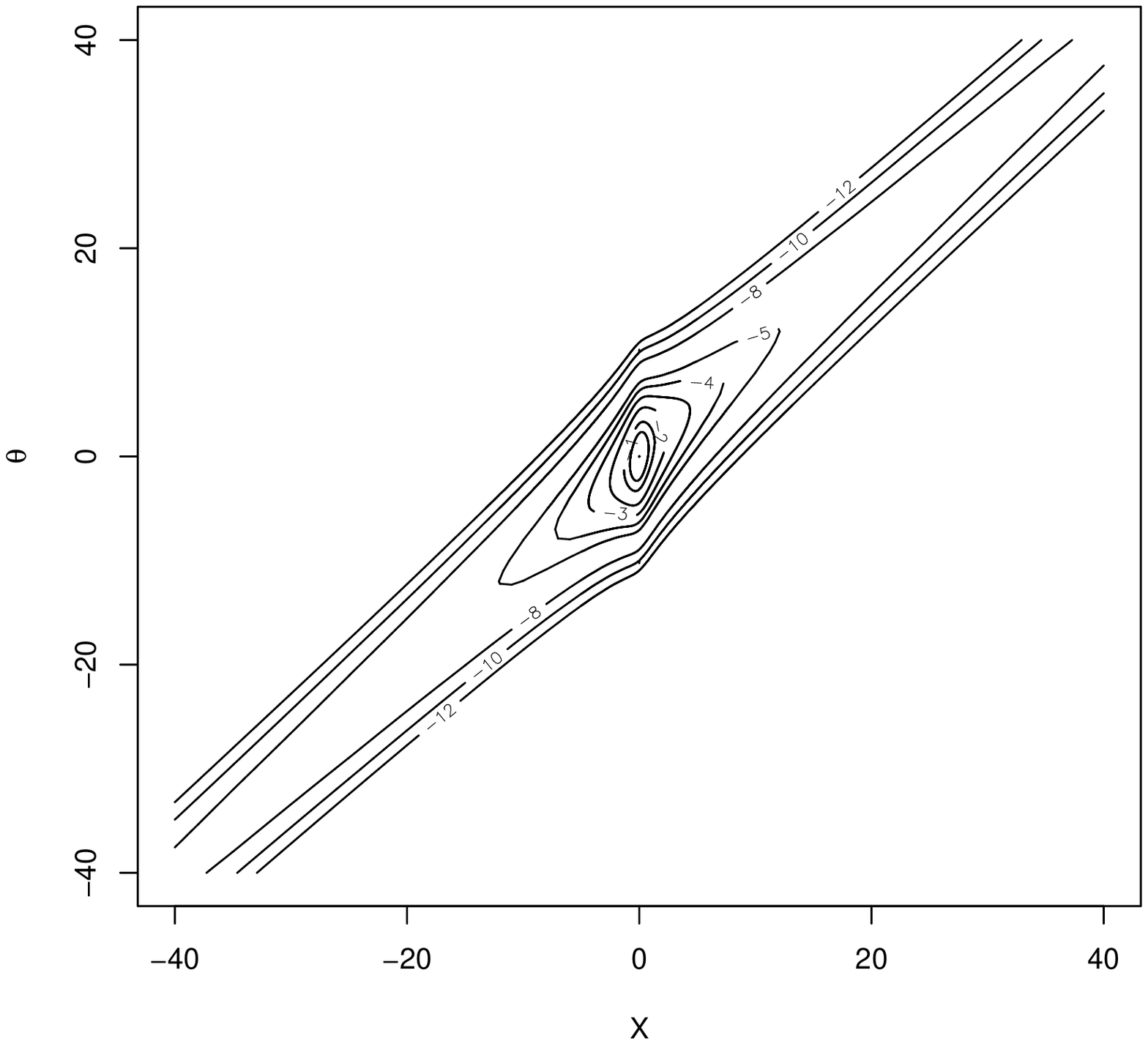,height=3.0in,width=3.0in, angle=270} \\
(a) & (b)
\end{tabular}
\label{f:cauchy-res} \caption{(a): two runs of the Gibbs sampler
under $\P0$ for the model (\ref{e:2line}) started at $\Theta_0=0$
(top) and $\Theta_0=200$ (bottom). (b): contours of the joint
posterior log-density  of $X$ and $\Theta$.}
\end{figure}
\end{center}

Consider a simplified version of (\ref{e:hier}) where
$m=m_1=C_1=D=1$,
\begin{eqnarray}
Y & = & X + Z_1 \nonumber \\
X & = & \Theta + Z_2. \label{e:2line}
\end{eqnarray}
Assume that $\L(Z_1)=\Ca(0,1)$, a standard Cauchy distribution,
$\L(Z_2)=\N(0,5)$, and $y=0$ is observed. Figure
\ref{f:cauchy-res}a shows the sampled values of $\Theta$ after two
independent runs of the Gibbs sampler, each of $10^4$ iterations.
The top one is started from the mode, $\Theta_0=0$, and
superficially it appears to be mixing well: the autocorrelation in
the series becomes negligible after 10 lags, and most convergence
diagnostic tests would assess that the chain has converged.
Nevertheless, the chain never exits the set $(-40,40)$, although
this is an event with stationary probability about $0.015$. The
second run, Figure \ref{f:cauchy-res}a bottom, is started from
$\Theta_0=200$, and the chain spends more than 4,000 iterations
wondering around $\Theta_0$. The contour plot of the joint
posterior log-density of $X$ and $\Theta$ in Figure
\ref{f:cauchy-res}b, provides an explanation: the contours look
roughly spherical near the mode, but they become asymptotically
concentrated around $x=\theta$ as $|\theta| \to \infty$. Thus,
restricted to an area around the mode, $X$ and $\Theta$ look
roughly independent, but in the tails they are highly dependent.
In fact, $\L(X - \theta \mid Y,\Theta=\theta) \to \N(0,5)$ as
$|\theta| \to \infty$, and we show in Section \ref{sec:comp} that
the Gibbs sampler which updates $X$ and $\Theta$ converges
sub-geometrically. In contrast, $\L(\tX \mid Y,\Theta=\theta) \to
\L(\tX)$, as $|\theta| \to \infty$, and as we show in Section
\ref{sec:comp} the Gibbs sampler which updates $\tX$ and $\Theta$
is uniformly ergodic.
%
%

\section{Convergence of the Gibbs sampler for linear hierarchical models}
\label{results}

Given the  parametrisation $\Pg = ({\bf U}, \Theta )$,  the
two-component Gibbs sampler simulates iteratively from  $\L({\bf
U} | {\bf Y},\Theta=\Theta_{n-1})$, and $\L(\Theta | {\bf Y},{\bf
U}={\bf U}_n)$, where $\Theta_0$ is a starting value and $n \geq
1$ denotes the iteration number. This algorithm generates a Markov
chain $\{({\bf U}_n,\Theta_n)\}$ with stationary distribution
$\L({\bf U},\Theta \mid {\bf Y})$. The marginal chain
$\{\Theta_n\}$
is also Markov and reversible with respect to $ \L(\Theta \mid
{\bf Y})$
(Lemma 3.1. of \citep{MR95d:62133}). Moreover, it can be shown
\citep{MR2002g:60108} that the convergence rate of the joint chain
coincides with the convergence rate of the marginal chain,
$\{\Theta_n\}$. Notice that this result does not hold for Gibbs
samplers which update more than two components. In the sequel, for
any random variables $W$ and $V$, and probability law $\mu$, we
will use the short-hand notation,
$$\L(V \mid W \sim \mu) := \int \L(V \mid W=w) \mu(dw).$$

We will consider the convergence of $\{\Theta_n\}$ through the
total variation norm, defined as
\begin{displaymath}
\label{e:tv} \|\L_{h}(\Theta_n \mid {\bf Y},\Theta_0)-\L(\Theta
|{\bf Y}) \| = \sup_{|g| \leq 1} |{\bf
    E}_{h}\{g(\Theta_n) \mid {\bf Y},\Theta_0\} - {\bf E}\{g(\Theta) |{\bf Y}\}|.
\end{displaymath}
$\L_{h}(\Theta_n \mid {\bf Y},\Theta_0)$ is the distribution of
the chain after $n$ steps started from $\Theta_0$, and  ${\bf
E}_{h}\{g(\Theta_n) \mid {\bf Y}, \Theta_0\}$ is the expected
value of a real bounded function $g$ with respect to this
distribution. $\L_{h}(\Theta_n \mid {\bf Y},\Theta_0)$ clearly
depends on the parametrisation ${\bf U}=h({\bf X},\Theta)$, since,
\begin{displaymath}
\L_{h}(\Theta_1 \mid {\bf Y},\Theta_0)=\L \mbox{\Large \{ }\Theta~
\mbox{\Large $\mid$ } {\bf Y},{\bf U}\sim\L({\bf U \mid
Y},\Theta=\Theta_0) \mbox{\Large \}}. \label{e:st}
\end{displaymath}
%
%
%
Under standard regularity conditions (Theorem 13.0.1 of \cite{MT})
the total variation norm converges to 0 as $n \to \infty$. We say
that $\{\Theta_n\}$ is {\em geometrically ergodic} when  there
exist an $r<1$ and some function $M(\cdot)$, such that
\begin{equation}
\label{e:ge} \|\L_{h}(\Theta_n \mid {\bf Y},\Theta_0)-\L(\Theta
|{\bf Y}) \| \leq M(\Theta_0) r^n.
\end{equation}
The smallest $r$ for which (\ref{e:ge}) holds, say $r_{h}$, is
known as the {\em
  rate of convergence} of $\{\Theta_n\}$.
   However, the actual distance from stationarity will in
  general depend on the starting point and this is represented by
  the term $M(\Theta_0)$ in (\ref{e:ge}).
  When $M(\cdot)$ is bounded above,  $\{\Theta_n\}$ is called {\em uniformly ergodic}.
Uniform
  ergodicity is a valuable property, since it ensures that the
  convergence of the chain does not depend critically on the initial value
  chosen. Whilst this does not guarantee rapid convergence, it
  ensures that the ``burn-in'' problem cannot become arbitrarily bad from
certain starting points.

Geometric ergodicity is a qualitative stability property, and
geometrically ergodic algorithms may still converge slowly and
give Monte Carlo estimates with high variance (for example when
$r_{h}\approx 1$). However, algorithms which fail to be
geometrically ergodic can lead to various undesirable properties,
including the break down of the central limit theorem for ergodic
average estimates.
In this  case the simulation can be unreliable and the drawn
samples might  poorly represent the target distribution.

To keep nomenclature simple we will identify a parametrisation
${\cal P}=({\bf U},\Theta)$ with the Gibbs sampler which updates
${\bf U}$ and $\Theta$. Thus, we say that a parametrisation ${\cal
P}$ is geometrically (respectively uniformly) ergodic, if the
Gibbs sampler implemented using this parametrisation is
geometrically (respectively uniformly) ergodic.

\subsection{Gaussian models}
\label{sec:gs}

The Gibbs sampler for the Gaussian linear model is geometrically
ergodic with rate given in \citep{MR98h:60100}. In the simplified
model (\ref{e:2line}) assume that $\L(Z_i)=N(0,
\sigma_i^2),i=1,2$, and define $\kappa = \sx^2 / (\sx^2 + \sy
^2)$.
Then, \citep{MR2003180} building on the results of
\citep{MR98h:60100} showed that, when $U=h(X,\Theta)=X-\rho
\Theta$,
\begin{equation}
\label{ratepnc} r_h:=r_{\rho } = {(\rho - (1 - \kappa ))^2 \over
\rho ^2 \kappa + (1- \rho )^2 (1-\kappa )} =
\{{\mbox{corr}}(U,\Theta \mid Y)\}^2
\end{equation}
which gives rise to the two special cases of interest, $r_0=
1-\kappa$, $r_1=\kappa$.
%
%
In this setting, the dependence between $U$ and $\Theta$ is
appropriately quantified by the correlation coefficient,  and
(\ref{ratepnc}) shows that the larger the correlation the worse
the convergence. Many refinements and generalizations of these
results can be found in \citep{MR98h:60100}, \citep{MR2003180} and
\citep{MR1731488}.
Notice that both $\P0$ and $\Pone$ are geometrically ergodic.
$\P0$ converges rapidly when the observation equation is ``more
precise'' than the hidden equation, that is $\sigma _1 <<\sigma
_2$, and it converges slowly when the hidden equation is
relatively precise. $\Pone$ converges rapidly when the hidden
equation is relatively more precise.
%

\subsection{General theory for linear hierarchical models}
\label{sec:gen-th}

This  section gives general results which can be used to
characterise the stability of the Gibbs sampler on linear
hierarchical models of the form (\ref{e:random}) where the $X_i$s
are univariate and $D=1$. Our results are valid when $m>1$ and
$m_i>1$ (see Remark 1 in page \pageref{pg:remarks}), however in
order to keep the notation simple we will work with the simplified
model (\ref{e:2line}), where all $Y,X$ and $\Theta$ are scalars.
$\L(Z_1)$ and $\L(Z_2)$ are arbitrary symmetric distributions with
continuous bounded everywhere positive densities, $f_1$ and $f_2$
respectively; common examples include the Gaussian, the Cauchy and
the double exponential. This section gives the general results,
while Section \ref{sec:comp} applies them to characterise the
convergence of the Gibbs sampler for (a broad class of) linear
non-Gaussian hierarchical models. Section \ref{sec:latent} deals
with extensions where the $X_i$s are vectors of dependent
variables, therefore covering state-space and spatial models.
Nevertheless, the results even for the more structured models
follow relatively easily from the results of this section. All
proofs are deferred to Section \ref{sec:proofs}.

We begin by  introducing a collection of  posterior robustness
concepts, which are related with the behaviour of the conditional
posterior distribution $\L(U \mid Y,\Theta=\theta)$ as $|\theta|
\to \infty$. All these concepts have statistical interpretations
 but they turn out to provide the
required mathematical conditions for characterising the stability
of the Gibbs sampler, as we show in Theorems \ref{th:unifnon},
\ref{th:non-g}
 and \ref{th:geo} below.

\begin{definition}
The parametrisation $\Pg=(U,\Theta)$ is called:
\begin{enumerate}
\item partially tight in parameter (PTIP),  if for all $y$, there
is some
  $k>0$ such that,
\begin{equation}
\label{ptip} \limsup_{|\theta | \to \infty } \Prob (|U|>k|Y=y,
\Theta=\theta ) <1,
\end{equation}
\item geometrically tight in parameter (GTIP),  if there exist
  positive constants, $a$, $b$ (independent of $\theta$)
such that for all $\theta $,
$$
\Prob (|U| > x|Y=y, \Theta=\theta ) \le a e^{-b x}.
$$
\end{enumerate}
\end{definition}
 GTIP not only implies that $\L(U \mid Y,\Theta=\theta)$ is a tight
family of distributions, but also that the tail probabilities are
bounded exponentially. (We recall that a family of distributions
on the real line, say $F_{\theta}$, indexed by a scalar $\theta$,
is called {\em tight} when $\lim_{k \to \infty} \sup_{\theta}
F_{\theta}([-k,k]^c)=0$.) Clearly, GTIP is much stronger condition
than PTIP. We consider also the following model robustness
concepts.

\begin{definition}
\label{def:dur}
We say that the linear hierarchical model (\ref{e:2line}) 
is
\begin{enumerate}
\item robust in parameter  (RIP),  if
$$
\lim_{|\theta |\to \infty } \L (X |Y=y, \Theta=\theta ) =
\L(Z_1+y),
$$
\item robust in data (RID), if
$$
\lim_{|\theta |\to \infty } \L (\tX |Y=y, \Theta=\theta ) =
\L(\tX),
$$
\item   data uniformly relevant (DUR), if there exist positive
constants $d$, $k$ such that for all $|\theta |>k$,
$$
|\E \{X|Y=y, \Theta=\theta \}| \leq  |\theta|-d,
$$
\ifdraft \marginpar{DUR ensures that the data are always relevant
and induce a bias in the conditional expectation. Notice that it
implies 2 inequalities, only one of which is though of interest,
the one side when $\theta>0$ and the other side when $\theta<0$;
for example when $\theta>0$ we have that $-2\theta-d
<E(X-\theta)<-d$, but only the $E(X-\theta)<-d$ is of interest}
\fi
\item  parameter uniformly relevant (PUR), if there exist positive
constants $d$, $k$ such that for all $|\theta |>k$,
$$
\mbox{sgn}(\theta ) \E \{X- y|Y=y, \Theta=\theta \} \geq d.
$$
\end{enumerate}
\end{definition}

These definitions characterise the hierarchical model according to
how inference for $X$ (conditionally on $\Theta=\theta$) is
affected by a large discrepancy between the data $y$ and the prior
guess $\theta$. When the model is RIP inference for  $X$ ignores
$\theta$, and it is symmetric around $y$. Conversely, when the
model is RID inference for $X$ ignores the data and becomes
symmetric around $\theta$. When the model is DUR (PUR) the data
(the parameter) always influences  the conditional expectation of
$X$.  Notice that when the model is RIP $\P0$ is PTIP (although
not necessarily GTIP), and when it is RID   $\Pone$ is PTIP. The
example in Section \ref{sec:example} describes a RID model. A
model can be both  DUR and PUR (for example the Gaussian linear
model).
\begin{theorem}
\label{th:unifnon}
Consider the linear hierarchical model (\ref{e:2line}) where the
error densities $f_1$ and $f_2$ are continuous, bounded and
everywhere positive. If $\P0$ ($\Pone$) is PTIP, then it is
uniformly ergodic.
\end{theorem}

\begin{theorem}
\label{th:non-g}
Consider the linear hierarchical model (\ref{e:2line}) where the
error densities $f_1$ and $f_2$ are continuous, bounded and
everywhere positive. If the model  is RID then $\P0$  is not
geometrically ergodic, and if the model is RIP then $\Pone$ is not
geometrically ergodic.
\end{theorem}
The proof Theorem \ref{th:non-g} is based on the general Theorem
\ref{th:rwlike} about Markov chains on the real line, which is
stated and proved in Section \ref{sec:proofs}.


\begin{theorem}
\label{th:geo} 1.\@ If  the model is DUR, $\Pone$ is GTIP, and
$\L(Z_2)$ has finite moment generating function in a neighbourhood
of $0$, then $\P0$ is geometrically ergodic.
\ifdraft \marginpar{note that when P1 is GTIP, the distribution of
|X-theta| has geometric tails and it concentrates around 0 for all
theta. If DUR then there is a skewness in the distribution} \fi
2.\@ If the model is PUR, $\P0$ is GTIP, and $\L(Z_1)$ has finite
moment generating function in a neighbourhood of $0$, then $\Pone$
is geometrically ergodic.
\end{theorem}
The theorems are proved by establishing a geometric drift
condition. The requirements of GTIP for $\Pone$ ($\P0$) and finite
moment generating function for $\L(Z_2)$ ($\L(Z_1)$) are in order
to tilt exponentially the linear drift condition provided by DUR
(PUR).

\subsection{Characterising the stability of the Gibbs
sampler according to the distribution tails of the error terms}
\label{sec:comp}

In this section, building upon the general theory of Section
\ref{sec:gen-th}, we characterise the stability of the Gibbs
sampler on the linear hierarchical model (\ref{e:2line})  for
different specifications of $\L(Z_1),\L(Z_2)$. Although we
consider the error distributions in Table \ref{tab:dists}, our
proofs remain valid for much broader families of distributions
(see Remark 2 on page \pageref{pg:remarks}).
\begin{table}
\begin{center}
\begin{tabular}{|c|c|c|}
\hline  Distribution & Code & Density $g(x)$ up to proportionality
\\ [0mm] \hline Cauchy & C & $\sigma ^2/ (1+x^2)$ \\ [0mm]
\hline  Double exponential & E & $\exp \left\{ -|x| / \sigma
\right\}$ \\ [0mm] \hline  Gaussian & G & $\exp \left\{
-(x/\sigma)^2 / 2 \right\}$
\\ [0mm] \hline  Exponential power distribution & L & $\exp \left\{
-{|x/\sigma|^{\beta } } \right\},\ \ \beta >2$ \\ [0mm] \hline
\end{tabular}
\end{center}
\caption{Distributions for the error terms and their densities. In
the paper they are coded according to the letter in the middle
column. } \label{tab:dists}
\end{table}
Notice that the exponential power distribution contains both the
Gaussian ($\beta=2$) and the double exponential ($\beta=1$) as
special cases. Here we consider densities with tails lighter than
Gaussian ($\beta>2$). For the use of this distribution  in
Bayesian robustness see \citep{MR2018034}.

We shall specify linear models giving first $\L(Z_1)$ and then
$\L(Z_2)$, for instance the $(C, E)$ model corresponds to
(\ref{e:2line}) with Cauchy distribution for $Z_1$, and  double
exponential distribution for $Z_2$. For each model we have two
parametrisations, thus two algorithms, $\P0$ and $\Pone$. When we
refer to the stability of an algorithm we shall write U, G, and N
to refer to  uniform, geometric and non-geometric (i.e.
sub-geometric) ergodicity, respectively.

\begin{theorem}
\label{th:tableresult} The stability $\P0$ and $\Pone$ is given in
Table \ref{tab:tailsgeom}.
\medskip
\begin{table}
\begin{center}
\begin{tabular}{cc}
\begin{tabular}{|c|c|cccc|}
\hline \multicolumn{6}{|c|}{{\bf
Stability of $\P0 $}}\\
\hline
  \multicolumn{6}{|c|}{\ \ \ \ \ \ \ \ \ \ \ \ \ \ \ \ \ \ $\L (Z_1)$ } \\
\hline
\multicolumn{2}{|c}{}& C & E & G & L \\
\hline
& C & U & U & U & U \\
$\L (Z_2)$ & E & N & G/U & U & U \\
& G & N & G & G & G \\
& L & N & G & G & G \\
\hline
\end{tabular}
&
\begin{tabular}{|c|c|cccc|}
\hline \multicolumn{6}{|c|}{{\bf
Stability of $\Pone $}}\\
\hline
  \multicolumn{6}{|c|}{\ \ \ \ \ \ \ \ \ \ \ \ \ \ \ \ \ \ $\L (Z_1)$ } \\
\hline
\multicolumn{2}{|c}{}& C & E & G & L \\
\hline
& C & U & N & N & N \\
$\L (Z_2)$ & E & U & U/G & G & G \\
& G & U & U & G & G \\
& L & U & U & G & G \\
\hline
\end{tabular}
\end{tabular}
\end{center}
\caption{Stability $\P0$ (left) and $\Pone$ (right) for the linear
hierarchical model  (\ref{e:2line}) for specifications of the
distribution of the error terms as in Table \ref{tab:dists}. }
\label{tab:tailsgeom}
\end{table}
\end{theorem}


\medskip
\noindent {\em Remark 1.}
The determining factor in classifying the stability of a
parametrisation is the tail behaviour of $\L(Z_1)$ and $\L(Z_2)$.
Thus, Theorem \ref{th:tableresult} generalises to the case of
multiple random effects and observations:
\begin{eqnarray*}
Y_{ij} & = &  X_i + Z_{1ij} ,~j=1,\ldots,m_i\\
X_i  & = &\Theta + Z_{2i},~i=1,\ldots, m
\end{eqnarray*}
where $Z_{1\cdot \cdot}$ and $Z_{2 \cdot }$ are independently
distributed identically to $\L(Z_1)$ and $\L(Z_2)$ respectively.
 This extension is immediate where
obvious sufficient statistics exist (the C and N cases). However,
since proving formally the full generalisation would be extremely
tedious (although in the same lines as in Section
\ref{sec:proofs}), we do not attempt it here. \label{pg:remarks}

\medskip
\noindent {\em Remark 2.} The same results can be obtained when
any of the distributions considered in Table \ref{tab:tailsgeom}
is replaced by another symmetric distribution with the same tail
behaviour, which possess a bounded continuous everywhere positive
density.

\medskip
\noindent {\em Remark 3.} Different results hold when a proper
prior for $\Theta $ is imposed. In this case the convergence
improves.

\medskip
\noindent {\em Remark 4.} The results of Theorem
\ref{th:tableresult}
 are independent of  the actual value of $y$.
This does not necessarily hold in  other contexts.

\medskip
\noindent {\em Remark 5.} In the $(E,E)$ model, the stability
depends on the ratio of the scale parameters in $\L(Z_1)$ and
$\L(Z_2)$. Depending on this ratio convergence can be either
geometric or uniform (see Section \ref{sec:proofs} for details).

\medskip
\noindent {\em Remark 6.} The following heuristic can be derived
from Table \ref{tab:tailsgeom}: convergence of $\P0$ is best when
$\L(Z_1)$ has  lighter tails  than $\L(Z_2)$,  and worst when it
has heavier tails.   The situation for $\Pone$ is the reverse.
Both algorithms become more stable the lighter the tails of
$\L(Z_1)$ and $\L(Z_2)$ become.

\subsection{Convergence of the grouped Gibbs sampler}
\label{sec:collapsed}

An alternative augmentation scheme and sampling algorithm can be
adopted when one of the error distributions, say $\L(Z_2)$ for
convenience, is Gaussian and the other, say $\L(Z_1)$,  is a scale
mixture of Gaussian distributions. Several symmetric distributions
belong in this class, for instance the Student-t (thus the Cauchy)
and the double exponential \citep{MR0359122}. In this case, $Z_1$
can be represented as $Z_1=V/Q$, where $V$ has a standard Gaussian
distribution and $Q$ is positive and independent of $V$. We can
treat  $Q$ as missing data and construct a {\em three-component}
Gibbs sampler which updates iteratively $X$, $Q$ and $\Theta$ from
their conditional distributions.  (When ${\bf X}=(X_1,\ldots,X_m)$
then ${\bf Q}=(Q_1,\ldots,Q_m)$ where $Q_i$ is independent from
$Q_j$ for every $i \ne j$). A major computational advantage of
this approach is that $\L(X \mid Y,\Theta,Q)$ is  Gaussian and it
can be easily sampled. Notice that $Q$ and $\Theta$ are
independent given $X$, thus we can implement the Gibbs sampler
using a {\em grouped} scheme \citep{MR1294740} where $\Theta$ and
$Q$ are updated in one block. It is of interest to know whether
the convergence of this grouped Gibbs sampler is better than the
convergence of the {\em collapsed} Gibbs sampler (as defined in
\citep{MR1294740}), where $Q$ has been integrated out. The
``Three-schemes Theorem'' of \citep{MR1294740} states that the
norm of the transition operator of the grouped Gibbs sampler is
larger than the one which corresponds to the collapsed Gibbs
sampler. This result, however, is not enough to guarantee that the
collapsed sampler will have better convergence rate.

In order to give a concrete answer, we consider the  important
special case, where $\L(Z_1)$ is the Cauchy distribution,
therefore $Q \sim \Ga(1/2,1/2)$. We have the following
proposition, whose proof is based on Theorem \ref{th:rwlike}.

\begin{proposition}
\label{lem:collapsed} The grouped Gibbs sampler is not
geometrically ergodic.
\end{proposition}
This result remains true for a number of random effects $m>1$, and
it will hold  for more general Student-t distributions. This
result has important practical implications especially in
algorithms for latent Gaussian models, considered in Section
\ref{sec:latent}. It is also significant that it contrasts the
result obtained by \citep{MR1345197}, who establishes geometric
ergodicity for variance component models (of which the model
considered here is a special case). However, the result in
\citep{MR1345197} is true when the number of data $Y_{ij}$, $m_i$,
per random effect $X_i$ is larger than some number bigger than
one, whereas in Lemma \ref{lem:collapsed} we take $m_i=1$.

\section{Latent Gaussian process models} \label{sec:latent}

In this section we consider a rather specific though useful model
and demonstrate that the results of Section \ref{sec:gen-th} can
be extended quite readily to this context giving some clear-cut
conclusions and advice for practical implementation. The results
below are certainly not the most general possible, but it is hoped
that the method of proof will indicate how analogous models might
be addressed.
\begin{theorem}
\label{th:geostat} Consider the latent Gaussian process model:
\begin{eqnarray}
\label{eq:lgp}
\bY  & = &  \bX + \bZ _1 \nonumber \\
\bX  & = &  \b1 \Theta +  {\bf \Sigma} ^{1/2} \bZ _2 \nonumber
\end{eqnarray}
where $\bZ_1 = \{Z_{11}, \ldots Z_{1p}\}$ is a vector of
independent and identically distributed standard Cauchy random
variables, $\bZ_2 = \{Z_{21}, \ldots Z_{2p}\}$ is a vector of
independent and identically distributed standard Gaussian random
variables, and $\b1 $ is a vector of $1$'s. ${\bf \Sigma} $ is
assumed known and a flat is prior is assigned to $\Theta$. Then
1.\@ $\P0$ fails to be geometrically ergodic; 2.\@ $\Pone$ is
uniformly ergodic.
\end{theorem}
As we remarked on page \pageref{pg:remarks}, the result holds when
the Cauchy is generalised to a Student-t with any degrees of
freedom. The MCMC for latent Gaussian process models  is often
implemented using a different augmentation scheme. As in Section
\ref{sec:collapsed}, we can augment the model with ${\bf
Q}=(Q_1,\ldots,Q_p)$, where $\L(Q_i)=\Ga(1/2,1/2)$. However, a
similar argument as in the proof of Proposition
\ref{lem:collapsed} shows that the Gibbs sampler which updates
${\bf X,Q}$ and $\Theta$ is not geometrically ergodic.

\begin{figure}[htb]
\label{fig:later-res}
\psfig{figure=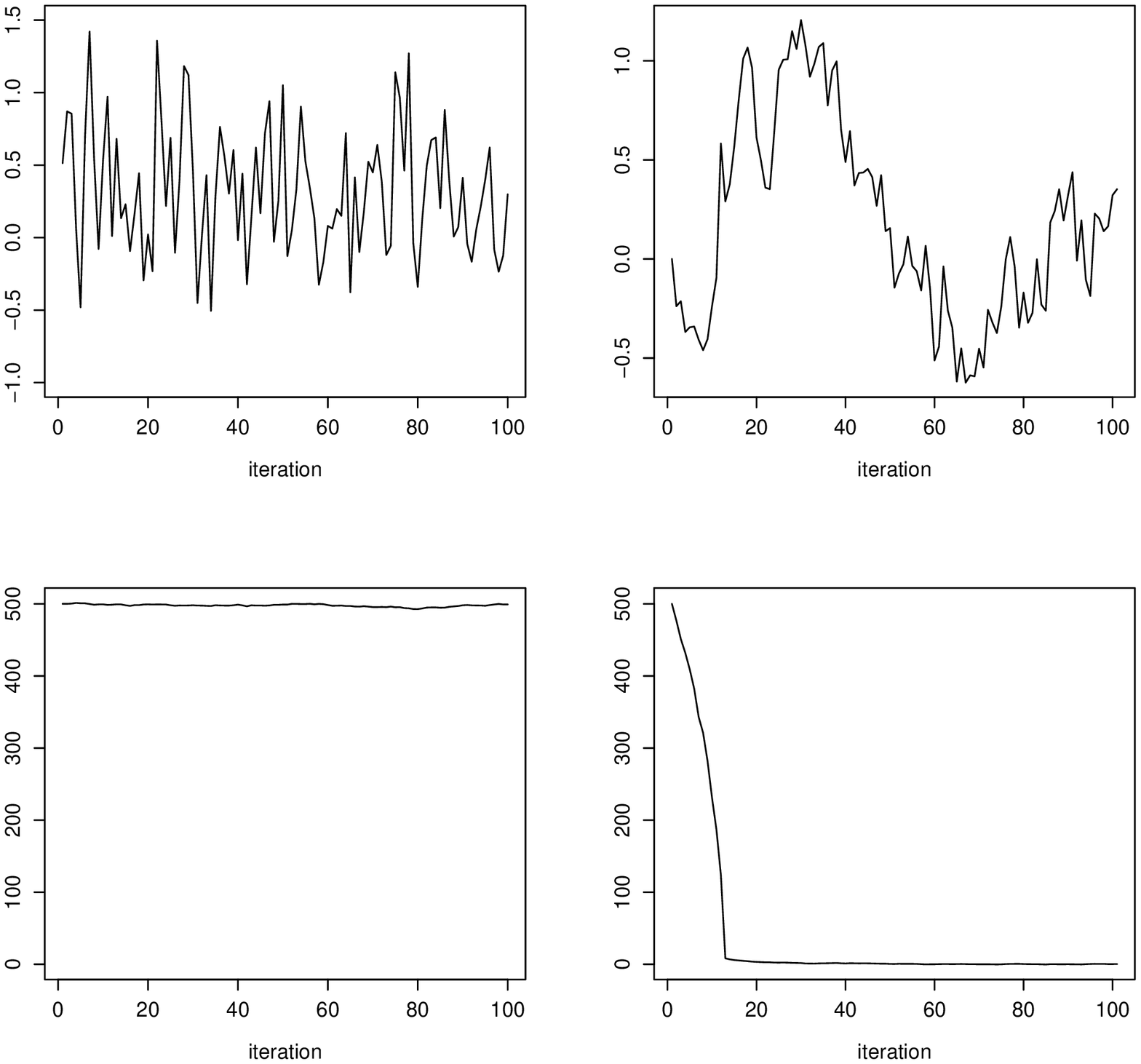,height=5.0in,width=3.0in,
angle=270} \caption{Two runs of $\P0$ (left) and $\Pone$ (right)
with two different starting values: $\Theta_0=0$ (top) and
$\Theta_0=500$ (bottom). }
\end{figure}

As a numerical illustration we consider a linear non-Gaussian
state-space model: $X_1,\ldots,X_p$ are consecutive draws from an
AR(1) model, which are observed with Cauchy error. We have
simulated $p=100$ data from this model using $\Theta=0$. The
update of $\Theta$ given $\bX$ is from a Gaussian distribution,
however the update of $\bX$ given $\Theta$ and $\bY$ is
non-trivial. We update all the states together using a highly
efficient Langevin algorithm, see \citep{robustMCMC} for details.
Moreover, we perform several updates of $\bX$ for every update of
$\Theta$ so that our results are not critically  affected by not
being able to simulate directly from $\L(\bX \mid \bY,\Theta)$.
Figure \ref{fig:later-res}  depicts our theoretical findings.
$\P0$ has a random walk-like behaviour in the tails, whereas
$\Pone$ returns rapidly to the modal area. On the other hand,
$\P0$ mixes better than $\Pone$ around the mode. Note that the
instability of $\P0$ in the tails is not due to lack of
information about $\Theta$ but due to the robustness properties of
the model.

\comment{
\begin{figure}
\label{fig:latent-dat}
\psfig{figure=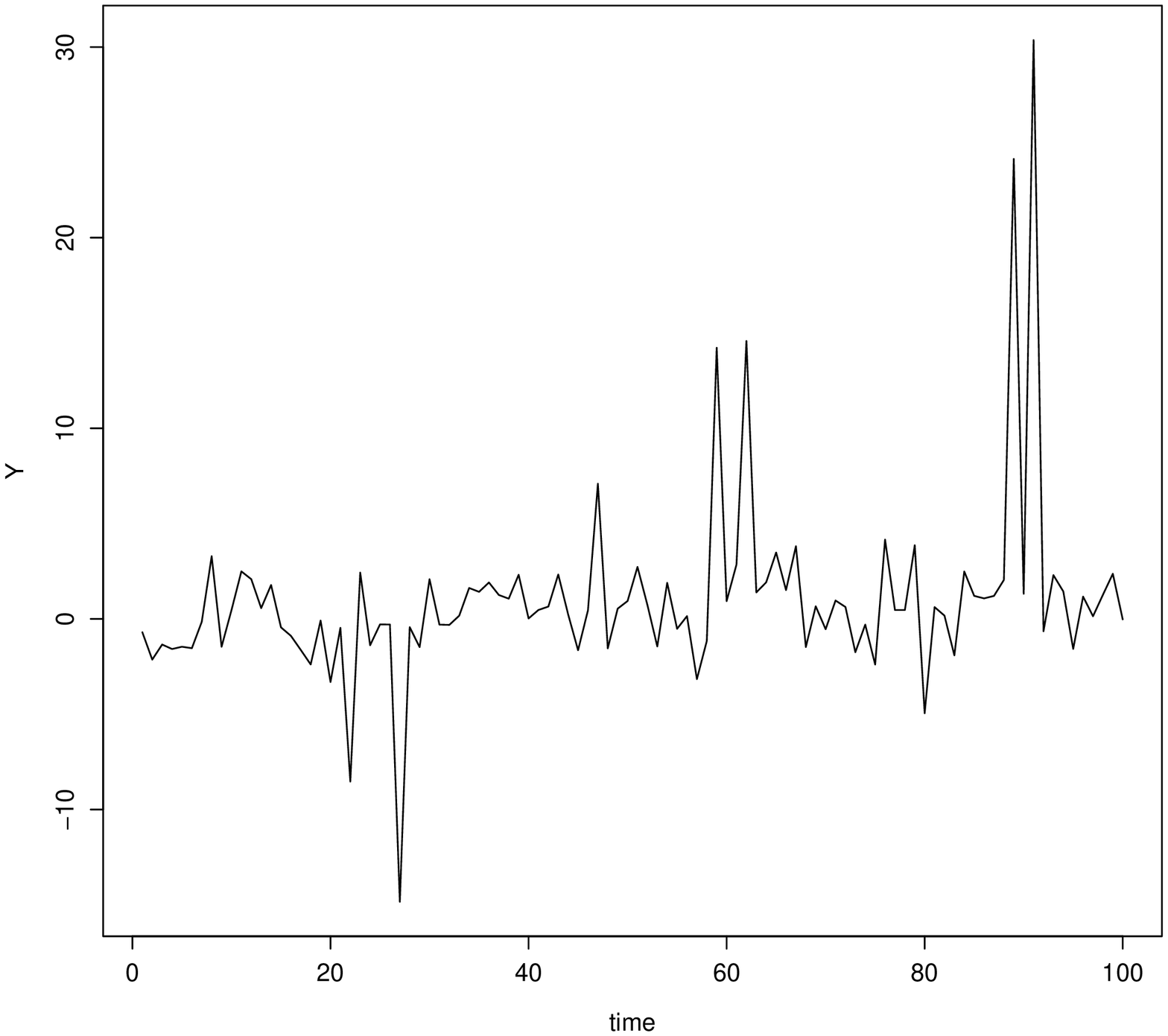,height=5.0in,width=3.0in,
angle=270} \caption{Data simulated from the state-space model
using $\Theta=0$. }
\end{figure}
}

 In this context it is definitely advisable to mix between $\P0$
 and $\Pone$, i.e to use a hybrid
sampler which at every iteration with some probability  updates
$(\Theta,{\bf X})$ and with the remaining probability it updates
$(\Theta,{\bf \tX})$. This hybrid sampler will inherit the uniform
ergodicity from $\Pone$ but it will also mix well around the modal
area.

\section{Discussion}
\label{sec:discuss}

We have obtained rigorous theoretical results for the stability of
the Gibbs sampler which explores the posterior distribution
arising from a broad class of linear hierarchical models. We have
also proved results regarding more complicated hierarchical models
with latent Gaussian processes, and we have compared different
sampling schemes. We have shown how the model structure dictates
which parametrisation should be adopted for improving the
convergence of the Gibbs sampler.

Our results are certainly not the most general possible, though
the method of proof we have used indicates clearly how analogous
problems might be addressed. As an example of this, it is easy to
extend the conclusions of Table \ref{tab:tailsgeom} to the case
where the light-tailed distributions are replaced by (say) uniform
distributions on finite ranges.  The robustness concepts of PTIP,
GTIP, RIP and RID are already stated in a general form, while the
concepts of DUR and PUR can be translated in a natural way using
Lyapunov drift conditions. Families of models to which we are
currently investigating extensions of our methods, include
stochastic volatility models prevalent in finance. This is the
subject of on-going research by the authors.

The general heuristic is clear - the stability of the centred and
non-centred algorithms, $\P0$ and $\Pone$ respectively, depends on
the relative tail behaviour of $\L(Z_1)$ and $\L(Z_2)$, with the
centred method being more stable when $\L(Z_1)$ is relatively
light tailed, and the non-centered being more stable when
$\L(Z_2)$ is relatively light tailed. An additional conclusion of
Table \ref{tab:tailsgeom} is that, as expected, both algorithms
possess comparatively more stable convergence properties the
lighter the tails  of $\L(Z_1)$ and $\L(Z_2)$ become.

The main message of the paper for the MCMC practitioner is a
positive one: the competition between $\P0$ and $\Pone$ works to
the user's benefit. Our results suggest that a combination of
$\P0$ and $\Pone$ is often desirable. When the tails of the error
distributions are very different we have found that one of the
algorithms might be very good for visiting the tails of the target
distribution whereas the other for exploring the modal area (as
for example we demonstrate in Figure \ref{fig:later-res}).
Therefore, it is advisable to use a hybrid Gibbs sampler which at
every iteration with some probability updates $(\Theta,X)$ and
with the remaining probability it updates $(\Theta,\tX$).
Moreover, by linking the stability of the Gibbs sampler to the
robustness properties of the hierarchical model we provide
intuition which can be found useful for models outside the scope
of this paper.

Another interesting product of this work is that linear
re-parametrisations, which can substantially improve the
convergence rate in (approximately) Gaussian models, might be of
little relevance when the tail behaviour of $\L(Z_1)$ is very
different from $\L(Z_2)$. For example, in (C,G) model, where the
observation error is Cauchy and the prior for $X$ is Gaussian, we
can prove that the Gibbs sampler which updates $U=X-\rho \Theta$
and $\Theta$ is sub-geometrically ergodic for all $\rho<1$,
whereas it is uniformly ergodic for $\rho=1$ as we already know
from Theorem \ref{th:tableresult}. This emphasizes the special
role of $\Pone$, which differs because of the prior independence
it induces on $\tX$ and $\Theta$. This result suggests that
conditional augmentation (as in \citep{MR1452025}) algorithms
might fail to be geometrically ergodic when $\P0$ does.

All the results presented here are specific to the Gibbs sampler,
however our findings are clearly relevant to contexts where
certain direct simulation  steps have to be replaced by
appropriate Metropolis-Hastings steps (as for example in the
simulation illustration in Section \ref{sec:latent}).

It is worth mentioning that once we have established geometric
ergodicity for an algorithm, it is important to obtain computable
bounds on the rate of convergence. We have not attempted to do so,
since it is outside the focus of this paper. For advances in this
direction see for example \citep{MR1345197,MR1888447}.

One interesting feature resulting from this paper is that the
marginal chain $\{\Theta_n\}$ of the Gibbs sampler on linear
non-Gaussian models often behaves asymptotically (i.e in the
tails) like a random auto-regression of the form:
$$
\Theta _n \ = \  \rho _n \Theta _{n-1} + \epsilon _n
$$
where $\rho _n $ is a random variable taking values in $[0,1]$,
and $\epsilon _n$ is an error term. For instance in the (G, G)
case of Theorem \ref{th:tableresult} for $\P0$  ($\Pone$) $\rho
_n$ is deterministically equal to $r_0$ ($r_1$) defined in Section
\ref{sec:gs}. The cases where we demonstrate that the algorithm is
random-walk like correspond to taking $\rho_n = 1$ ({\em almost
surely}). Furthermore in a number of cases, $\rho _n$ is genuinely
random. For instance, in the (E, E) case with identical rates,
$\rho _n \sim U[0,1]$. In the (C,C) case, we find that $\rho _n $
takes the value $0$ or $1$ with probabilities determined by the
scale parameters of the Cauchy distributions involved.

An extension of our ideas is possible for hierarchical models with
more levels. For instance consider the linear structure given by
\begin{eqnarray}
Y  & = & \Theta _1 + Z_1  \nonumber \\
\Theta _{i} & =  & \Theta_{i+1} + Z_{i+1}, \ \ \ i = 1, \ldots d-1
\ ,
 \label{e:long-hier}
\end{eqnarray}
with a flat prior on $\Theta _d$. Since $Y$ is the only
information available, the posterior tails of $\Theta _1, \Theta
_2 \ldots $ become progressively heavier. If at any stage, $Z_i$
has lighter tails than $Z_{i-1}$, then whenever $\Theta _{i-1}$
and $\Theta _{i+1}$ strongly disagree, the conditional
distribution of $\Theta _i$ given $Y, {\bf \Theta} _{-i}$ will
virtually ignore $\Theta _{i-1}$ and hence the data. This will
lead to potential instabilities in the chain in components $\Theta
_i, \Theta _{i+1}, \ldots , \Theta _d$. We call this phenomenon
the {\em quicksand} principle, and this is the subject of ongoing
investigation by the authors.

\section{Proofs of main results}
\label{sec:proofs}

In the sequel we will use $\pi$ to denote the density of any
stationary measure, in particular $\pi(\theta \mid y)$ and $\pi(x
\mid y,\theta)$ will be the Lebesgue densities of $\L(\Theta \mid
Y=y)$ and $\L(X \mid Y=y,\Theta=\theta)$ respectively. With
$p(\cdot,\cdot)$ we denote the transition density of a Markov
chain, and with $\Theta_0$ and $\Theta_1$ the consecutive values
of the marginal chain $\{\Theta_n\}$.

{\sc proof of Theorem \ref{th:unifnon}}. We show the result for
 $\P0$, since the corresponding result for $\Pone$ can be proved in an analogous
 way. In particular, we show that when $\P0$ is PTIP, the
 transition density of the the marginal chain $\{\Theta_n\}$, is
 such that $\inf_{\theta_0} p(\theta_0,\theta_1)>0$, and $p$ is also
 continuous in $\theta_1$. This guarantees
 uniform ergodicity by Theorem 16.0.2  of
\citep{MT}.
 \begin{eqnarray*}
p(\theta_0,\theta_1)  & = & \int f_2(|x-\theta_1|) \pi(x \mid
y,\theta_0)dx \geq \int_{-k}^k f_2(|x-\theta_1|) \pi(x \mid
y,\theta_0)dx \\
& \geq & \inf_{|x| \leq k} f_2(|x-\theta_1 |) ~\Prob (|X|
 \leq k | Y=y, \Theta=\theta_0 ),
 \end{eqnarray*}
for $k$ such that (\ref{ptip}) holds. Since $f_1$ and $f_2$ are
everywhere positive, bounded and continuous, $\Prob (|X|
 \leq k | Y=y, \Theta=\theta_0 )$ is also positive and continuous in $\theta_0$, therefore
 by the PTIP property it follows that $\inf_{\theta_0} \Prob (|X| \leq
k | Y=y, \Theta=\theta_0 )>0$. Moreover, $\inf_{|x| \leq k}
f_2(|x-\theta_1 |)$, is positive  and  continuous in $\theta_1$,
thus the result follows. \qed

The proof of Theorem \ref{th:non-g} requires Theorem
\ref{th:rwlike}, hence it is proved on page \pageref{pg:3.4}.
 The proof of Theorem \ref{th:geo} requires  the following lemmas.
\begin{lemma}
\label{lem:drift}
\begin{enumerate}
\item If (\ref{e:2line}) is DUR and the parametrisation $(\tX,
\Theta )$ is GTIP, then for all sufficiently small $\alpha>0$,
\begin{eqnarray*}
\E \left \{e^{\alpha X} | Y, \Theta=\theta \right \} & \leq &
e^{\alpha
\theta} (1-\alpha d/2),~~~\mbox{for}~ \theta> k \\
\E \left \{e^{-\alpha X} | Y, \Theta=\theta \right \} & \leq &
e^{-\alpha \theta} (1-\alpha d/2),~~~\mbox{for}~ \theta< -k,
\end{eqnarray*}
where $k,d$ are defined in Definition \ref{def:dur}. \item If
(\ref{e:2line}) is PUR and the parametrisation $(X, \Theta )$ is
GTIP, then for all sufficiently small $\alpha>0$,
\begin{eqnarray*}
\E \left \{e^{\alpha (y-\tX)} | Y=y, \Theta=\theta \right \} &
\leq & e^{\alpha
\theta} (1-\alpha d/2),~~~\mbox{for}~ \theta> k \\
\E \left \{e^{-\alpha (y-\tX)} | Y=y, \Theta=\theta \right \} &
\leq & e^{-\alpha \theta} (1-\alpha d/2),~~~\mbox{for}~ \theta<
-k,
\end{eqnarray*}
\end{enumerate}
\end{lemma}

\proof 1. We will prove only the first inequality, for $\theta>k$,
since the other is proved in a similar fashion. We define
$G_{\theta }(t ) = \E \left\{ e^{t~(X -\theta) } \mid
Y,\Theta=\theta \right \}$, which is finite for all sufficiently
small $t>0$, say $0<t<t_0$ for some $t_0$, and for all $\theta$,
since by the GTIP assumption $\L(|X-\theta| \mid Y,\Theta=\theta)$
has exponential or lighter tails. By a second order Taylor series
expansion of $G_{\theta }(t )$ around $t=0$, we obtain for some
$0<t _1<
 t_0$, and for $\theta >k$,
\begin{eqnarray*}
G_\theta (t )  &  = &  1+t~ \E\{X-\theta \mid Y,\Theta=\theta \}+
{t ^2 \over 2}~ \E\left\{ (X-\theta)^2
 e^{t_1~ (X-\theta)} | Y, \Theta=\theta \right\} \\
& \leq &
 1 - td  + {t ^2 \over 2}  \E\left\{ (X-\theta)^2
 e^{t_1 (X-\theta)} | Y, \Theta=\theta \right\}.
\end{eqnarray*}
Now pick $\alpha <  t_1$  small enough so that for all $\theta>k$
$\alpha \E\left\{ (X-\theta)^2
 e^{t_1 (X-\theta)} | Y, \Theta=\theta \right\}<d$. Such $\alpha$ exists due
 to the GTIP assumption. Then,
$ G_\theta (\alpha )  \leq 1 - \alpha d /2$, and the result
follows.  2. It is proved as 1, recognising that $\tX=X-\theta$.
\qed \ifdraft \marginpar{what we basically need to prove here is
that if for a random variable U we know that E(U)<0 and that its
distribution has geometric tails, then G(t):=E(exp(tU))<1 for
sufficiently small t. This is intuitive, since the geometric tails
ensure the existence of exponential moments for small t, G(0)=1
and G'(0)<0. If G is continuous we are done} \fi

\begin{lemma}
\label{lem:drift-abs}  \begin{enumerate} \item If (\ref{e:2line})
is DUR and the parametrisation $(\tX, \Theta )$ is GTIP, then for
all sufficiently small $\alpha>0$,
\begin{eqnarray*}
\E \left \{e^{\alpha |X|} | Y, \Theta=\theta \right \} & \leq &
e^{\alpha |\theta|} (1-\alpha d/2)+K,~~~\mbox{for}~ |\theta|> k,
\end{eqnarray*}
where $k,d$ are defined in Definition \ref{def:dur}, and
$0<K<\infty$. \item If (\ref{e:2line}) is PUR and the
parametrisation $(X, \Theta )$ is GTIP, then for all sufficiently
small $\alpha>0$,
\begin{eqnarray*}
\E \left \{e^{\alpha |y-\tX|} | Y=y, \Theta=\theta \right \} &
\leq & e^{\alpha |\theta|} (1-\alpha d/2)+K,~~~\mbox{for}~
|\theta|> k,
\end{eqnarray*}
where $k,d$ are defined in Definition \ref{def:dur}, and
$0<K<\infty$.
\end{enumerate}
\end{lemma}
\proof  1. We prove the result for $\theta>0$ exploiting the first
inequality given in Lemma \ref{lem:drift}. The case $\theta<0$ is
proved analogously but exploiting the second inequality of Lemma
\ref{lem:drift}. Notice that
$$\E \left \{e^{\alpha |X|} | Y, \Theta=\theta
\right \} \leq \E \left \{e^{\alpha X} | Y, \Theta=\theta \right
\} + \int_{-\infty}^{0} e^{-\alpha x} \pi(x\mid y,\theta)dx, $$
thus, due to Lemma \ref{lem:drift} we only need to show that the
second term of the sum above can be bounded above for all
$\theta$. Recall $a,b$ from the GTIP Definition \ref{def:dur}.
Choose $\alpha <b$. Using integration by parts, we find that the
second summand is bounded above by,
$ e^{-b\theta}[a+\alpha/(b-\alpha)]$, which can easily be bounded
above for all $\theta>k$. 2. It is proved as 1,  recognising that
$\tX=X-\Theta$. \qed

{\sc proof of Theorem \ref{th:geo}} 1. We prove the result
establishing a geometric drift condition for the marginal chain
$\{\Theta_n\}$, using the function $V(\theta)=e^{\alpha
|\theta|}$, for appropriately chosen $\alpha > 0$.  Notice first
that $\L(\Theta \mid Y,X=x) \equiv \L(\Theta \mid X=x)$ is
symmetric around $x$ and has a finite moment generating function
in a neighbourhood of the origin. Thus, working as in  Lemma
\ref{lem:drift} and Lemma \ref{lem:drift-abs}, we can show that
for all sufficiently small $\alpha>0$, there exists $K_1>0$ and
$\epsilon>0$, such that,
$$
\E\{ e^{\alpha |\Theta|} \mid X=x \} \leq \left (1+\alpha^2
\epsilon \right )e^{\alpha |x|}+K_1.
$$
Then, for $|\theta_0|>k$, and appropriate $K_1>0,K>0$,
\begin{eqnarray*}
\E \{e^{\alpha |\Theta_1|} \mid Y,\Theta_0=\theta_0  \} & = & \E
\{ \E\{
e^{\alpha |\Theta_1|} \mid X_1\} \mid Y,\Theta_0=\theta_0 \} \\
& \leq & \E \{(1+\alpha^2 \epsilon)e^{\alpha |X_1|}+K_1  \mid
Y,\Theta_0=\theta_0 \} 
\\
& \leq & (1+\alpha^2\epsilon)(1-\alpha d /2) e^{\alpha
|\theta_0|}+K 
\\
& \leq & (1-\alpha \delta) e^{\alpha |\theta_0|}+K.
\end{eqnarray*}
Now since standard arguments (see for example \citep{robtwebo})
show that compact sets are small for this problem, the Gibbs
sampler is shown to be geometrically ergodic by Theorem 15.0.1 of
\citep{MT}.

2. The second result is proved almost identically. Notice that
$\L(\Theta \mid Y=y,\tX=x)$ is symmetric around $y-x$ and
possesses finite moment generating function in a neighbourhood of
0, thus as we showed above, for all sufficiently small $\alpha>0$,
there exists a $K_1>0$ such that,
$$
\E\{ e^{\alpha |\Theta|} \mid Y=y,\tX=x \} \leq \left (1+\alpha^2
\epsilon \right )e^{\alpha |y-x|}+K_1.
$$
Using Lemma \ref{lem:drift-abs} and arguing as in 1 proves the
theorem. \qed

Before proving Theorems \ref{th:non-g} and \ref{th:tableresult} we
need the following general result about Markov chains on the real
line.
\begin{theorem}
\label{th:rwlike} Let $\{W_n\}$ be an ergodic and reversible with
respect to a density $\pi $, Markov chain  on ${\bf R}$ with
transition density $p(x, y)$ which is  random walk-like in the
tails, in the sense that there is a continuous
 positive symmetric density $q$ such that
\begin{equation}
\lim_{|x|\to \infty }p(x, x+z) = q(z),~z \in {\bf R}.
\label{eq:enumerate2}
\end{equation}
Then
\begin{enumerate}
\item $\pi$ has heavy tails, in the sense that
\begin{equation}
\label{eq:rwlikecvgce} \lim_{x \to \infty } {\log \int_x^{\infty}
\pi(u)du \over x} = \lim_{x \to  \infty } {\log
\int_{-\infty}^{-x} \pi (u)du \over -x } = 0\ ;
\end{equation}
\item $\{W_n\}$ is not geometrically ergodic.
\end{enumerate}
\end{theorem}
{\sc proof} 1. We will prove the result for $x \to \infty$, since
the case $x \to - \infty$, is proved in the same way. Fix $z,
\delta \in {\bf R}^+$, and let $W$ denote a random variable which
has density $\pi$.  By (\ref{eq:enumerate2}), there exists $k>0$
such that for $x>k$
$$
{ p(x + z, x) \over p(x, x+z) } \le (1+\delta )\ .
$$
This uses the fact that $q(z) >0$. Thus by reversibility, and for
$x
> k$,
$$
{ \pi (x) \over \pi (x+z) } = { p(x + z, x) \over p(x, x+z) } \le
(1+\delta )\ ,
$$ so that
\begin{equation}
\label{e:pibd} \pi (x +z) \ge (1+\delta )^{-1} \pi (x)\ .
\end{equation}
Integrating (\ref{e:pibd}) over $x >k$, gives that
\begin{equation}
\label{e:prob-heavy} \Prob (W> k+z) \ge (1+\delta )^{-1} \Prob (W>
k)\ .
\end{equation}
Iterating this expression, and after some algebra, we get that
$$
\lim_{n \to \infty} {\log \Prob(W>k+nz) \over n} \geq -\delta,
$$
which, since $\delta$ can be chosen arbitrarily small, proves the
statement.

2. The second follows from the following standard capacitance
argument; see  \citep{robtwebo} for similar arguments for MCMC
algorithms and \citep{lawsok} for an introduction to Cheeger's
inequality using capacitance. Cheeger's inequality for reversible
Markov chains implies that geometric ergodicity must fail if we
can find $k>0$, such that the probability
$$
\Prob \left(| W_1| \leq k \mid   W_0 \sim \pi_{(-k,k)^c} \right)
$$
is arbitrarily small, where we use $\pi_{(-k,k)^c}$ to denote the
density $\pi$ restricted and re-normalised to the set $\{|x|>k
\}$. Notice that (\ref{e:prob-heavy}) implies that for
sufficiently large $k$, for $|x|
>k$, and any $l>0$, there
$$
\Prob (|W_1 | > x + l | W_0 >k ) \ge (1+ \delta )^{-1}\ge 1-\delta
\ .
$$
Now choose $l$ sufficiently large that $\int_l^\infty q(u) du
<\delta $ then for all $|x| > k$,
$$
 \Prob\left(|W_1|<k \right)
\le \Prob(|W_1|<k \mid W_0 \sim \pi_{(-k,k)^c}) + \Prob (
|W_1-W_0|>l)
$$
which converges as $|x| \to \infty $ to a limit bounded by $3
\delta$. Since $\delta $ is arbitrary, the result is  proved. \qed

{\sc proof of Theorem \ref{th:non-g}} \label{pg:3.4} we prove the
theorem for the case where the model is RID, since the proof when
the model is RIP
 is identical. We will show that under the assumptions the
marginal chain $\{\Theta_n\}$ generated by the centred Gibbs
sampler is random walk-like, thus by Theorem \ref{th:rwlike} $\P0$
is not geometrically ergodic. By assumption, $\lim_{|\theta |\to
\infty } \L (\tX |Y, \Theta=\theta ) = \L(\tX)$, which is
symmetric around 0, and let $F$ denote its corresponding
distribution function. Therefore $\Prob(X \leq \theta+z \mid
Y,\Theta=\theta) \to F(z)$, as $|\theta| \to \infty$.
Notice that,
$$p(\theta_0,\theta_0+z) = \int f_2(|x-\theta_0-z|) dF(x \mid
Y,\Theta=\theta_0) = \int f_2(|u-z |) dF(u+\theta_0 \mid
Y,\Theta=\theta_0),$$
therefore, since $f_2$ is bounded, $p(\theta_0,\theta_0+z) \to
\int f_2(|u-z|) dF(u)=q(z)$, as $|\theta_0| \to \infty$, where $q$
is a symmetric density around  $0$.\qed

{\sc proof of Theorem \ref{th:tableresult}}  Throughout the proof
we shall use the following notation: $f_1$  and $f_2$ denote the
density of $Z_1$ and $Z_2$ respectively (at least up to
proportionality), and we define
\begin{eqnarray*}
f_\theta (x)  & = & f_1(|y-x|) f_2(|x-\theta| ), \end{eqnarray*}
thus, $ \pi(x \mid y,\theta)  =  f_\theta (x)/c_\theta$,  where
$c_\theta $   is the normalisation constant. Any scale parameter
involved in $f_i$ will be denoted by $\sigma_i,~i=1,2$.

For each model, we first prove the result for $\P0$ and
subsequently for $\Pone$. We will prove the statements
corresponding to the upper triangular elements of the $\P0$ and
$\Pone$ tables. This is without loss of generality, since we can
write (\ref{e:2line}) as
\begin{eqnarray*}
{\tilde X} & = & Y-\Theta -Z_1   \\  {\tilde X} & = &  Z_2  \ .
\end{eqnarray*}
Since the actual value of $Y$ does not affect convergence (as can
be verified by our proofs below), we may as well set it to be $0$,
and since $\L(Z_1),\L(Z_2)$ are symmetric around 0, the model
written above under a non-centred parametrisation coincides with
(\ref{e:2line}) under a centred parametrisation but with the error
distributions interchanged. We first prove the results concerning
the diagonal elements.

\medskip
\noindent {\bf The $(C, C)$ model} \newline \noindent We prove the
result by verifying the PTIP property. The result then follows by
Theorem \ref{th:unifnon}. Notice that in this model, $c_\theta =
\int_{-\infty } ^{\infty } f_\theta (x) dx = 2 \int _{-\infty }
^{(y+\theta)/2} f_\theta (x) dx $.  We show that $\P0 $ is PTIP by
demonstrating that for arbitrary $k>0$,
$$
\liminf_{|\theta |\to \infty } \int_{y-k}^{y+k} f_\theta
(x)/c_\theta~ dx
>0\ .
$$
By symmetry, it is enough to prove this statement for large
positive $\theta $ values, so from now on we shall assume that
$\theta > y$.

For $x < (y + \theta )/2$, $1+ (y-\theta )^2 \le 1+4(x-\theta)^2
\le 4(1 + (x-\theta )^2)$,  so that $ c_\theta \le  4 / \pi (1 +
(y-\theta )^2)$. Moreover, notice that when $x \in (y-k,y+k)$,
then there exist a $d>0$ (depending on $k,y$), such that for all
$\theta>d$,
$$
{1+(y-\theta)^2 \over 1+(x-\theta)^2} \geq {1+(y-\theta)^2 \over
1+(y+k-\theta)^2 } \geq 1/2.
$$
Therefore, for $\theta>d$,
\begin{eqnarray*}
\int_{y-k}^{y+k} f_\theta (x)/c_\theta~ dx  & \geq &
\int_{y-k}^{y+k}
{ 1+(y-\theta)^2 \over 4\pi(1+(y-x)^2)(1+(x-\theta)^2) } dx \\
& \geq & {1 \over 8} \int_{y-k}^{y+k} {1 \over \pi(1+(y-x)^2)}>0,
\end{eqnarray*}
which proves the result. The result for $\Pone$ is proved
identically.

\medskip
\noindent {\bf The $(E,E)$ model} \newline \noindent Without loss
of generality we assume that $f_1(x) \propto \exp\{-|x| \}$, and
$f_2(x) \propto \exp \{-|x|/\sigma  \}, \sigma>0$. The stability
of the Gibbs sampler depends on whether $\sigma <1$,  $\sigma = 1$
or $\sigma >1$, thus we consider these cases separately. Again by
symmetry it is enough to consider $y < \theta $.
\begin{enumerate}
\item \ $\sigma = 1$: here we can write
$$
f_\theta (x) = \left \{ \begin{array}{ll}  {1 \over 4} e^{2x - y -
\theta }, & x < y  \\ {1 \over 4} e^{-(\theta -y )}, & y \le x \le
\theta  \\ {1 \over 4} e^{ y + \theta -2x }, & x > \theta \,.
\end{array} \right.
$$
From this it is easy to demonstrate that $E(\Theta _1|\Theta
_0=\theta_0)=(y+\theta_0 )/2$. Since all compact sets are small
for the Markov chain $\{\Theta _n\}$ this is enough to demonstrate
geometric ergodicity by Theorem 15.0.1 of \citep{MT}. \item \
$\sigma >1$: \ here we can write:
$$
f_\theta (x) = \left \{ \begin{array}{ll}  {1 \over 4} e^{(1+
\sigma)x - y - \sigma \theta }, & x < y  \\ {1 \over 4} e^{y
-\sigma \theta +(\sigma -1 ) x}, & y \le x \le \theta  \\ {1 \over
4} e^{ y + \sigma \theta -(1+ \sigma )x }, & x > \theta \,.
\end{array} \right.
$$
Direct algebra shows that
$$
\E\{ X-\theta \mid Y, \Theta=\theta \} =
p_1(\theta)(Y-1)+[p_2(\theta)+p_3(\theta)-1]\theta +p_2(\theta)r
(\theta) +{p_3(\theta) \over \sigma+1}-{p_2(\theta) \over \sigma
-1},
$$
where $p_1(\theta)+p_2(\theta)+p_3(\theta)=1$, and as $\theta \to
\infty$, $p_2(\theta) \to (\sigma+1)/(2 \sigma),p_1(\theta) \to 0,
r(\theta) \to 0$. Therefore,
$$
\lim_{\theta \to \infty } \E \{X -\theta|Y, \Theta=\theta \}  \le
{-2 \over \sigma ^2 -1}\ ,
$$
and the model is DUR. Since $\Pone $ is easily seen to be GTIP, by
part 1 of Theorem \ref{th:geo},  $\P0 $ is geometrically ergodic.
\item \ $\sigma < 1$: \ Here, in an analogous way to the above, we
can demonstrate that $\P0 $ is RIP therefore by Theorem
\ref{th:unifnon}, $\P0 $ is uniformly ergodic.

Due to symmetry, the results for $\Pone$ are proved in a similar
fashion, notice however, that $\Pone$ is uniformly ergodic when
$\sigma>1$.
\end{enumerate}

\medskip
\noindent {\bf The $(G,G)$ model} \newline \noindent This is
covered in \citep{MR98h:60100,MR2003180} and reviewed in Section
\ref{sec:gs}.

\medskip
\noindent {\bf The $(L,L)$ model} \newline \noindent We assume
that $f_1(x) \propto\exp\{ - |x/\sigma_1|^\beta\}$, $f_2(x)\propto
\exp\{ - |x/\sigma_2|^\beta\}$, and we  let $a = \beta /(\beta
-1)$. Again by symmetry we just consider the case $y < \theta $.
For large $\theta $, $\L(X | Y, \Theta=\theta) $ converges weakly
and in $L^1$ to a point mass at $ \rho \theta + (1-\rho )y $ where
$$
\rho = {\sigma _1^{-a}  \over \sigma _2 ^{-a}  + \sigma _1^{-a} }.
$$
As a result, neither $\P0$ nor $\Pone$ are GTIP, so it is not
possible to establish geometric ergodicity using the DUR and PUR
properties (which hold for this model) in conjunction with Theorem
\ref{th:geo}. Instead, we have to construct directly a geometric
drift condition. However, this is rather easy. Notice that since
$\L(\Theta \mid X=x)$ is symmetric around $x$, we can find a $b>0$
such that $\E\{|\Theta| \mid X=x  \} \leq |x|+b$. Moreover, for
any $\epsilon >0$, there is some $k>0$, such that for all
$\theta|>k$, $\E\{|X-y| \mid Y=y,\Theta=\theta\} \leq
(1+\epsilon)\rho |\theta-y|$, thus
$$
\E\{|\Theta _1 - y | \ | \Theta _0=\theta_0\} \le b + \rho (1 +
\epsilon ) | \theta _0 -y|
$$
which implies geometric ergodicity for $\P0$ since compact sets
can easily be seen to be small. The result for $\Pone$ is proved
identically.

\medskip
\noindent {\bf The $(C,G),(E,C)$ and $(L,C)$  models}
\newline \noindent We show that the model is RIP, therefore since
$\P0$ is PTIP,  by Theorem \ref{th:unifnon} $\P0$ is uniformly
ergodic, and by Theorem \ref{th:non-g} $\Pone$ is not
geometrically ergodic. Notice, however, that for any $x$, using
dominated convergence we can show that $c_\theta/f_2(|x-\theta|)
\to 1$, as $|\theta| \to \infty$. The argument is that, for any
$u$, $f_2(|u-\theta|)/f_2(|x-\theta|) \to 1$, and the ratio is
bounded above (as a function of $\theta$) by a function of $u$
which is integrable with respect to $f_1$, as long as $f_1$ has
exponential tails or lighter, which is the case in the models
considered here. However, since $f_{\theta}/c_\theta \to
f_1(|y-x|)$, and this limit is a proper density, it follows that
the corresponding distribution functions converge and $\L(X \mid
Y=y,\Theta=\theta) \to \L(|Z_1-y|)$ as $|\theta| \to \infty$.

\medskip
\noindent {\bf The $(G,E)$ model}\newline \noindent Calculations
show that
$$
\lim_{\theta \to \infty } \L (X | Y,\Theta=\theta  ) = N ( y +
\sigma_1^2/\sigma_2, \sigma_1^2),~\textrm{and}~ \lim_{\theta \to
-\infty } \L (X | Y,\Theta=\theta  ) = N ( y -
\sigma_1^2/\sigma_2, \sigma_1^2),
$$
therefore $\P0$ is PTIP (but not RIP) and by Theorem
\ref{th:unifnon} uniformly ergodic. The above result, however,
shows that the model is PUR, and since all conditions of Theorem
\ref{th:geo} are satisfied, $\Pone$ is geometrically ergodic.

\medskip
\noindent{\bf The $(L,E)$ model}\newline \noindent The result is
proved as above.

\medskip
\noindent {\bf The $(L,G)$ model}\newline \noindent Here (perhaps
surprisingly) $\P0$ is not PTIP but the model is  DUR and PUR, and
both $\P0$ and $\Pone$ are GTIP so that Theorem \ref{th:geo} can
be applied.

\qed

\medskip
\medskip

{\sc proof of Lemma \ref{lem:collapsed}} Consider the Gibbs
sampler with initial value $X_0$ which updates $(\Theta,Q)$ first
and then $X$. Direct calculation gives that $\L(Q \mid
Y=y,X=x,\Theta=\theta)=\Ga(1,(y-x)^2/2)$, $\L(X \mid
Y=y,\Theta=\theta,Q=q) = N(\theta/(q+1)+qy/(q+1),1/(q+1))$,
therefore $\L(X_1-X_0 \mid Y=y,Q_1=q) =
N(q(y-X_0)/(q+1),1+1/(q+1))$. However, since $q \to 0$ in
probability, when $X_0 \to \infty$, the algorithm is random
walk-like in the tails and by Theorem \ref{th:rwlike} fails to be
geometrically ergodic. \qed

\medskip
\medskip

{\sc proof of Theorem \ref{th:geostat}}  It is easy to demonstrate
that the model is RID,

$$ \lim_{|\theta |\to \infty }\L ({\bf \tX}  | \bY,\Theta=\theta) =
N_p \left(\bz ,\Sigma \right).$$ Therefore, $\Pone$ is PTIP and by
Theorem \ref{th:unifnon} is uniformly ergodic. Since
$$
\Theta | \bX \sim \left({\b1 \Sigma ^{-1} \bX \b1 \over \b1 \Sigma
^{-1} \b1}, {1 \over \b1 \Sigma ^{-1} \b1} \right)
$$
this implies that for the Gibbs sampler using $\P0$,
$$
\lim_{|\theta _n |\to \infty } \L (\Theta _{n+1}-\theta_n | \Theta
_n=\theta_n ) = N\left( 0, {2 \over \b1 \Sigma ^{-1} \b1} \right)\
,
$$
Therefore by Theorem \ref{th:rwlike}, geometric ergodicity fails.

\bibliographystyle{plain}

\bibliography{stability}

\end{document}

*****************************************